\documentclass{elsart}
\usepackage{graphicx}
\usepackage{amssymb}
\begin{document}

\begin{frontmatter}
\title{Analysis of the low-energy $\pi^\pm p$ differential cross sections of the CHAOS Collaboration}
\author[EM]{E. Matsinos{$^*$}},
\author[GR]{G. Rasche},
\address[EM]{Centre for Applied Mathematics and Physics, Zurich University of Applied Sciences, Technikumstrasse 9, P.O. Box, CH-8401 Winterthur, Switzerland}
\address[GR]{Institut f\"{u}r Theoretische Physik der Universit\"{a}t, Winterthurerstrasse 190, CH-8057 Z\"{u}rich, Switzerland}

\begin{abstract}
This paper presents the results of an analysis of the low-energy $\pi^\pm p$ differential cross sections, acquired by the CHAOS Collaboration at TRIUMF \cite{chaos,denz}. We first analyse separately the $\pi^+ p$ and the $\pi^- p$ 
elastic-scattering measurements on the basis of standard low-energy parameterisations of the $s$- and $p$-wave $K$-matrix elements. After the removal of the outliers, we subject the truncated $\pi^\pm p$ elastic-scattering databases 
into a common optimisation scheme using the ETH model \cite{glmbg}; the optimisation failed to produce reasonable values for the model parameters. We conclude that the problems we have encountered in the analysis of these data are 
due to the shape of the angular distributions of their $\pi^+ p$ differential cross sections.
\end{abstract}
{$^*$}{Corresponding author. E-mail: evangelos.matsinos@zhaw.ch, evangelos.matsinos@sunrise.ch; Tel.: +41 58 9347882; Fax: +41 58 9357306}
\end{frontmatter}

\section{\label{sec:Introduction}Introduction}

This is the second of three papers addressing issues of the pion-nucleon ($\pi N$) interaction at low energies (pion laboratory kinetic energy $T \leq 100$ MeV). The goal in this study is to investigate the self-consistency of the 
$\pi^\pm p$ elastic-scattering differential cross sections (DCSs) of Refs.~\cite{chaos,denz} (in accordance with our naming convention, we hereafter refer to these data as DENZ04); these measurements, which had been acquired at TRIUMF 
in 1999 and 2000, have not been included in our last two phase-shift analyses (PSAs) \cite{mworg,mrw1} (to be referred to as UZH06 and ZUAS12, respectively). In the case of UZH06, we did not notice that the measurements were already 
available for some time~\footnote{The $546$ DENZ04 DCS values have been given in tabulated form in Ref.~\cite{denz}, and were available two years prior to the main publication of the CHAOS Collaboration \cite{chaos}.}. In the case of 
ZUAS12, we made a conscious decision to avoid modifying our UZH06 database prior to the assessment of the self-consistency of any candidate additions, considering in particular the amount of the DENZ04 data which almost matches the 
size of our UZH06 database. In the present work, we will give arguments supporting our position not to use the DENZ04 data and to retain our initial UZH06 $\pi^\pm p$ elastic-scattering databases for future use.

We will analyse the DENZ04 data as if it comprised the entire $\pi^\pm p$ elastic-scattering database at low energies; this way, a possible failure when testing the degree of applicability of our PSA to these data cannot be blamed on 
other experimental data. We will follow the method introduced in Section 4 of Ref.~\cite{mworg} and developed to its current form in Ref.~\cite{mrw1} (see Section 2 therein). We will first investigate the self-consistency of the $\pi^+ p$ 
measurements on the basis of suitable low-energy parameterisations of the $s$- and $p$-wave $K$-matrix elements; the most recent values of any constants, which are used in the parameterisation of these quantities, may be found in 
Ref.~\cite{mrw1}. Any outliers will be removed from the data, one data point at a time, until the data sets are consistent and ready for further analysis. At the next step, the $\pi^- p$ elastic-scattering measurements will be analysed. 
After removing any outliers also from these data, we will investigate the possibility of analysing both reactions in a common optimisation scheme; in that part of the study, we will use both the low-energy parameterisations of the $s$- 
and $p$-wave $K$-matrix elements and, finally, the ETH model \cite{glmbg}.

The last part of the study will be dedicated to the reproduction of the absolute normalisation of the DENZ04 data on the basis of our ZUAS12 solution. We will show that the ZUAS12 solution, based on the bulk of our established $\pi^+ p$ 
database at low energies, is incompatible with the shape of the angular distribution of the DENZ04 $\pi^+ p$ differential cross sections.

\section{\label{sec:Method}Method}

The determination of the observables from the hadronic phase shifts has been given in detail in Section 2 of Ref.~\cite{mworg}. For $\pi^+ p$ scattering, one obtains the partial-wave amplitudes from Eq.~(1) of that paper and determines 
the no-spin-flip and spin-flip amplitudes via Eqs.~(2) and (3). Finally, the observables are evaluated from these amplitudes via Eqs.~(13) and (14). For $\pi^- p$ elastic scattering, the observables are determined on the basis of Eqs.~(15-20).

All the details on the analysis method (i.e., on the minimisation function, on the definitions of the scale factors, etc.) may be found in Section 2.2 of Ref.~\cite{mrw1}. The contribution $\chi_j^2$ of the $j^{th}$ data set to the overall 
$\chi^2$ is given therein by Eq.~(1). The scale factors $z_j$, which minimise each $\chi_j^2$, are evaluated using Eq.~(2); the minimal $\chi_j^2$ value for each data set (denoted by $(\chi_j^2)_{min}$) is given in Eq.~(3) and the scaling 
contribution (of the $j^{th}$ data set) to $(\chi_j^2)_{min}$ in Eq.~(4). Finally, the scale factors for free floating $\hat{z}_j$ (which we will use in Section \ref{sec:Reproduction}, when investigating the absolute normalisation of the 
DENZ04 data sets using the ZUAS12 solution as reference) are obtained via Eq.~(5); their total uncertainty $\Delta \hat{z}_j$ has been defined at the end of Section 2.2 of Ref.~\cite{mrw1}.

One statistical test will be performed for each data set, the one involving its contribution $(\chi_j^2)_{min}$ to the overall $\chi^2$. The corresponding p-value will be evaluated on the basis of $(\chi_j^2)_{min}$ and number of degrees 
of freedom of the data set (hereafter, the acronym DOF will stand for `degree(s) of freedom', whereas NDF will denote the `number of DOF'); for a data set with $N_j$ data points (none of which is an outlier), NDF is equal to $N_j$. The 
p-value for each data set will be compared to the confidence level $\mathrm{p}_{min}$ for the acceptance of the null hypothesis (implying no statistically-significant effects). The value of $\mathrm{p}_{min}$ is fixed to the equivalent 
of a $2.5 \sigma$ effect in the normal distribution, corresponding to about $1.24 \cdot 10^{-2}$.

To facilitate the repetitive use of the full description of the databases, we adhere to the following notation: DB$_+$ for the $\pi^+ p$ database; DB$_-$ for the $\pi^- p$ elastic-scattering database; DB$_{+/-}$ for the combined $\pi^\pm p$ 
elastic-scattering databases.

\section{\label{sec:Database}The DENZ04 data}

\subsection{\label{sec:General}General comments on the data}

The DENZ04 DB$_+$ consists of $275$ data points, acquired at five energies between $19.90$ and $43.30$ MeV. Two sets of values are available at $43.30$ MeV: one was taken at the same conditions as the data at the lower energies, whereas 
another was obtained with the target rotated by $64^\circ$; similarly to the notation in Ref.~\cite{denz}, we will identify the data set corresponding to the rotated target via the label `(rot.)'. Technically, the DENZ04 $\pi^+ p$ data 
must be assigned to only $6$ data sets \cite{chaos}. However, the measurements have been assigned to a total of $17$ data sets in the SAID database \cite{abws}, after splitting each data set (except at $19.90$ MeV) into three parts: 
forward-angle ($\theta < 35^\circ$), medium-angle ($35 < \theta \lesssim 150^\circ$), and backward-angle ($\theta \gtrsim 150^\circ$); $\theta$ denotes the centre-of-mass (CM) scattering angle. The $19.90$ MeV data do not cover scattering 
angles above $98.45^\circ$; therefore, the original data set has been split into two segments.

Some justification for this splitting of the original data sets may be found in Ref.~\cite{abws}, and rests on the variation of the event-selection algorithm with the scattering-angle interval. The essential point is that a coincidence 
measurement (i.e., the simultaneous detection of both the scattered pion and proton), enabling the vertex reconstruction at `medium' scattering-angle values, cannot be performed in the cases of very-forward and very-backward scattering 
at low energies; as a result, only the scattered pion had been detected in very-forward scattering, and only the scattered proton in very-backward scattering. Although we have analysed the DENZ04 data also the way the CHAOS Collaboration 
appears to suggest (i.e., by assigning the measurements for each reaction to only $6$ data sets), the results of our analysis clearly favour the splitting of the data sets into segments the way this is done in the SAID database. We will 
give the results of the optimisation only in the case of the split $\pi^+ p$ data sets. The assignment of the measurements to $17$ data sets enables the determination of the scale factors $z_j$ from data which are more `localised' (in 
terms of the scattering angle) and, as such, it implies a more favourable treatment of the data. Evidently, the appearance of any problems in case of the assignment of the data to $17$ data sets can only be exacerbated if only $6$ data 
sets are used.

The DENZ04 DB$_-$ comprises $271$ data points, taken at the same five beam energies of the DENZ04 DB$_+$; similarly to $\pi^+ p$, two sets of DCS values have been acquired at $43.30$ MeV. In SAID, the measurements have been assigned to 
$12$ data sets; we will do the same in the optimisation phase.

The final normalisation uncertainties reported by the CHAOS Collaboration \cite{chaos} are as follows: $5 \%$ at the three lowest energies and $7 \%$ for the $43.30$ MeV data sets. Asymmetric uncertainties have been given for the $37.10$ 
MeV data sets ($+5, -9 \%$); unable to treat asymmetric uncertainties in our software, we have decided to use the normalisation uncertainty of $7 \%$ (average of the two absolute values) for the $37.10$ MeV data~\footnote{At the time the 
present work is submitted for reviewing, the SAID group are still using the normalisation uncertainties as they appeared in the captions of the tables of Appendix B of Ref.~\cite{denz}, instead of those quoted in the final publication of 
the CHAOS Collaboration \cite{chaos}.}.

The SAID output for the DENZ04 $\pi^+ p$ data also contains the contribution of these data sets to their overall $\chi^2$ value. According to their results, the contribution of the $274$ data points~\footnote{The SAID group have excluded 
one data point belonging to the $37.10$ MeV medium-angle data set.} of the DENZ04 $\pi^+ p$ data to the overall $\chi^2$ is very large ($703.22$ units). Furthermore, the $\chi^2$ value for the DENZ04 $\pi^- p$ elastic-scattering data is only 
slightly better ($503.30$ for $271$ data points). On the basis of these numbers, it is evident that the WI08 phase-shift solution yields a statistically-poor reproduction of the DENZ04 data.

\subsection{\label{sec:K-Matrix_pi+p}Fits to the DENZ04 DB$_+$ using the $K$-matrix parameterisations}

The parameterisation of the $s$- and $p$-wave $K$-matrix elements for the low-energy $\pi^+ p$ scattering may be found in Section 3.1 of Ref.~\cite{mrw1}. The optimal values of the corresponding seven parameters ($\tilde{a}_{0+}^{3/2}$, 
$b_3$, $c_3$, $d_{33}$, $e_{33}$, $d_{31}$, and $e_{31}$) are obtained via the minimisation of the $\chi^2$ function (see Section 2.2 of Ref.~\cite{mrw1}). We will apply the same acceptance criteria to the DENZ04 measurements which were 
applied to the data in the ZUAS12 PSA.

The results of the optimisation procedure are shown in Table \ref{tab:Pi+p}. Since seven parameters are used to generate the fitted values, the NDF in the first fit to the DENZ04 DB$_+$ was $268$; the minimum value of $\chi^2$ was $486.9$. 
For the truncated DENZ04 DB$_+$, the minimum value of $\chi^2$ was $401.2$ for $260$ DOF in the fit. The values of the seven parameters of the fit came out far from those obtained in the fits to the 
truncated DB$_+$ of Ref.~\cite{mrw1}. The details on the truncated DENZ04 DB$_+$, as obtained from the final fit, are given in Table \ref{tab:DBpi+p}.

\subsection{\label{sec:K-Matrix_pi-p}Fits to the DENZ04 DB$_-$ using the $K$-matrix parameterisations}

The $I=3/2$ amplitudes were fixed from the final fit to the truncated DENZ04 DB$_+$ and were imported into the analysis of the DENZ04 DB$_-$. The parameterisation of the $s$- and $p$-wave $I=1/2$ $K$-matrix elements, suitable for the low-energy 
$\pi^- p$ elastic scattering, may be found in Section 3.2 of Ref.~\cite{mrw1}. Seven new parameters ($\tilde{a}_{0+}^{1/2}$, $b_1$, $c_1$, $d_{13}$, $e_{13}$, $d_{11}$, and $e_{11}$) are introduced at this stage. We present the steps in the process 
of removing outliers from the DENZ04 DB$_-$ in Table \ref{tab:Pi-p}; only three data points had to be removed. The final result for $\tilde{a}^{cc}$, obtained from the data, was around $0.041\:\mu_c^{-1}$, a value which differs by a factor of $2$ 
from the result $\tilde{a}^{cc}=0.0803(11)\: \mu_c^{-1}$ of Ref.~\cite{mrw1} ($\mu_c$ denotes the mass of the charged pion); the $\tilde{a}^{cc}$ value of Ref.~\cite{mrw1} roughly agrees with the result obtained from the pionic-hydrogen data. 
We traced this discrepancy to the very unusual final values of the parameters entering the modelling of the $s$- and $p$-wave $K$-matrix elements for the $\pi^+ p$ reaction (i.e., the values which fixed the $I=3/2$ amplitudes in the case of the fits 
to the DENZ04 DB$_-$). The details on each data set of the truncated DENZ04 DB$_-$, as obtained from the final fit, are given in Table \ref{tab:DBpi-p}.

\subsection{\label{sec:K-Matrix}Common fit to the DENZ04 DB$_{+/-}$ using the $K$-matrix parameterisations}

In order to give the two elastic-scattering reactions equal weight, we multiplied $(\chi^2_j)_{min}$ for each $\pi^+ p$ data set by
\[
w_+=\frac{N\!_+ + N\!_-}{2N\!_+}
\]
and for each $\pi^- p$ elastic-scattering data set by
\[
w_-=\frac{N\!_+ + N\!_-}{2N\!_-} \, ,
\]
where $N\!_+$ and $N\!_-$ represent the NDF in the two databases; we then added these quantities for all the data sets to obtain the overall $\chi^2$ value. The application of these `global' weights for the two reactions was made as a matter of 
principle; given the proximity of the $N\!_+$ and $N\!_-$ values in the case of the DENZ04 data, the effect of this weighting on our results is very small.

The common fit to the truncated DENZ04 DB$_{+/-}$ was subsequently performed, using $14$ parameters. This step was taken in order to examine whether any additional points (or data sets) had to be removed; none were identified. The common fit to 
the data yielded a $\chi^2$ value of $751.5$ for $521$ DOF in the fit.

\subsection{\label{sec:Model}Common fit to the truncated DENZ04 DB$_{+/-}$ using the ETH model}

So far in this paper, we have used standard low-energy parameterisations of the $\pi N$ amplitudes in terms of the pion CM kinetic energy. We will now use the ETH model which is based on Feynman diagrams. Details on the model, as well as on its seven 
parameters ($G_\sigma$, $K_\sigma$, $G_\rho$, $K_\rho$, $g_{\pi NN}$, $g_{\pi N \Delta}$, and $Z$) may be obtained from Refs.~\cite{mworg,mrw1}. This model was introduced in Ref.~\cite{glmbg} and was developed to its final form by the mid 1990s.

\subsubsection{\label{sec:ModelResults}Results}

The common fit of the ETH model to the truncated DENZ04 DB$_{+/-}$ yielded a $\chi^2$ value of $783.3$ for $528$ DOF in the fit. All results for the parameters of the ETH model turned out to be far from their `established' values. At the same time, 
the numerical evaluation of the correlation (Hessian) matrix failed and the positivity had to be enforced by the MINUIT software library \cite{jms} (which we exclusively used in the optimisation); hence, the uncertainties of the fit parameters could 
not be obtained.

Our results for the seven model parameters have shown good stability over the years, from the period when the fits were performed to old, outdated phase shifts to the present times when the fits are made directly to the contents of the low-energy 
$\pi N$ database. The database itself has also changed significantly over the last two decades. In any case, it is fair to say that, no matter which data were fitted to, the results for the model parameters always came out within a reasonable 
interval of values; this had been the case even when obvious outliers (e.g., the measurements of Ref.~\cite{brt}) were included in our database (e.g., see Ref.~\cite{m}).

The results for the model parameters, obtained from the common fit to the truncated DENZ04 DB$_{+/-}$, are very odd. Evidently, for whichever reasons, the parameters drift away from their `established' values, to unreasonable (or even unphysical) ones. 
It is therefore meaningless to give `optimal' values for the model parameters.

Despite the drift of the model parameters in the fit, we decided to determine the $s$- and $p$-wave phase shifts with the model-parameter values obtained in the fit to the truncated DENZ04 DB$_{+/-}$. The values of the DENZ04-based hadronic phase 
shifts (and their overall tendency with increasing energy) were found hard to accept. The final results for these phase shifts were far from the values established in Refs.~\cite{mworg,mrw1,abws}.

\subsubsection{\label{sec:Reproduction}Reproduction of the DENZ04 data on the basis of the ZUAS12 solution}

Given all these problems, we decided to investigate the reproduction of the DENZ04 data on the basis of the ZUAS12 prediction. Our goal in this part of the study is to identify the kinematical region(s) in which the DENZ04 data are poorly reproduced; 
if successful, we could pinpoint the origin of the problems we have encountered in the analysis of these measurements.

The DENZ04 measurements, normalised to the corresponding ZUAS12 predictions, are shown in Figs.~\ref{fig:PIPPE} and \ref{fig:PIMPE}; the outliers detailed in Tables \ref{tab:Pi+p} and \ref{tab:Pi-p} are also contained in these figures. Evidently, the 
angular distribution of the DB$_+$ disagrees with the shape obtained from the rest of the $\pi^+ p$ measurements in Ref.~\cite{mrw1}; on the other hand, the angular distribution (and the absolute normalisation) of the DB$_-$ is in reasonable agreement 
with the results of Ref.~\cite{mrw1}.

We will start with the reproduction of the measurements when the data sets are characterised only by the target configuration and the energy of the incident beam (i.e., following the suggestion of the CHAOS Collaboration \cite{chaos}). The results 
are shown in Table \ref{tab:Reproduction1}. We notice that the overall $\chi^2$ values of the reproduction (i.e., the sums of the corresponding $(\chi_j^2)_{min}$ values given in the table for the two elastic-scattering reactions) are: $547.7$ for 
$\pi^- p$ elastic scattering ($271$ data points) and $2446.0$ for $\pi^+ p$ ($275$ data points).

The reproduction of the DENZ04 measurements after the data sets have been split into $29$ segments in total (see Section \ref{sec:General}) are given in Table \ref{tab:Reproduction2}. We observe that the overall $\chi^2$ values of the reproduction drop 
for both reactions: to $506.5$ for $\pi^- p$ elastic scattering, to $747.3$ for $\pi^+ p$. The dramatic decrease in the latter case indicates that the problems we have encountered in the analysis of the DENZ04 data are mainly due to the shape of the 
$\pi^+ p$ angular distributions. The decrease in the case of the $\pi^- p$ elastic-scattering data (i.e., `unsplit' versus split data sets) is very moderate, indicating considerably fewer problems with the DENZ04 DB$_-$. The results after removing the 
outliers, detailed in Tables \ref{tab:Pi+p} and \ref{tab:Pi-p}, are shown in Table \ref{tab:Reproduction3}; the $\chi^2$ values drop further to $469.1$ and $665.7$ for $\pi^- p$ and $\pi^+ p$ elastic scattering, respectively. The scale factors for free 
floating $\hat{z}_j$, corresponding to the optimal reproduction of the absolute normalisation of the DENZ04 data on the basis of the ZUAS12 solution are given in Figs.~\ref{fig:sfpip} and \ref{fig:sfpim}, separately for the two elastic-scattering reactions. 
For $\pi^+ p$ scattering, three $\hat{z}_j$ values per energy are obtained (corresponding to the three angular intervals into which the measurements have been split, i.e., forward, medium, and backward angles); as earlier mentioned, the $19.90$ MeV data 
set does not cover backward angles. For $\pi^- p$ elastic scattering, two $\hat{z}_j$ values per energy are obtained (corresponding to the two angular intervals of the measurements, i.e., forward and medium/backward angles).

We recollect that the $\chi^2$ results of Ref.~\cite{mrw1} (for $\mathrm{p}_{min} \approx 1.24 \cdot 10^{-2}$) for the two reactions were: $371.0$ and $427.2$ for $321$ and $333$ DOF in the fit, for $\pi^- p$ and $\pi^+ p$ elastic scattering, respectively. 
The $F$-test performed on the two $\pi^- p$ elastic-scattering databases (i.e., on the truncated DENZ04 DB$_-$ and on the truncated ZUAS12 DB$_-$) results in the score value of $1.515$ for $268$ and $321$ DOF, corresponding to the p-value of about $1.9 \cdot 10^{-4}$. 
On the other hand, the $F$-test performed on the two corresponding $\pi^+ p$ databases results in the score value of $1.944$ for $267$ and $333$ DOF, resulting in a p-value of about $4.8 \cdot 10^{-9}$. These two results are sufficient to substantiate 
our position that the DENZ04 measurements are not compatible with the rest of the low-energy $\pi N$ database, as it emerged in our PSA of Ref.~\cite{mrw1}. In view of these striking differences, it makes no sense to include even part of the DENZ04 data, 
as they currently stand, in our PSAs.

The inspection of Fig.~\ref{fig:sfpip} shows that the extracted scale factors $\hat{z}_j$ for the DENZ04 DB$_+$ scatter to the extent that no coherent picture may be obtained from these results. A closer look, however, at the five entries for backward 
scattering demonstrates that these data can be reproduced well by our ZUAS12 solution~\footnote{Note that the backward-angle $25.80$ MeV data set had to be freely floated, when the self-consistency of the DENZ04 DB$_+$ was investigated using the $K$-matrix 
parameterisations, e.g., see Section \ref{sec:K-Matrix_pi+p}, as well as Tables \ref{tab:Pi+p} and \ref{tab:DBpi+p}; therefore, the normalisation of this data set is questionable even when the data set is compared only to the rest of the DENZ04 $\pi^+ p$ 
measurements. This implies that the seemingly-poor reproduction of the absolute normalisation of this data set on the basis of the ZUAS12 solution is less problematic than it appears to be. Indeed, if the scale factor of $0.8150$ of Table \ref{tab:DBpi+p} 
is applied to this data set, its resulting absolute normalisation will be compatible with our ZUAS12 solution.}. However, all scale factors which are obtained at forward and medium angles (except at $37.10$ MeV) show that the experimental data systematically 
exceed the `theoretical' values obtained on the basis of the optimal parameters of Ref.~\cite{mrw1}. The discrepancies reach the $35 \%$ level, with an average around $15 \%$.

On the other hand, the scale factors for free floating $\hat{z}_j$ obtained in the case of the truncated DENZ04 DB$_-$ (Fig.~\ref{fig:sfpim}) cluster well around the expectation value of $1$. Therefore, the absolute normalisation of the DENZ04 DB$_-$ 
appears to be compatible with the results obtained in Ref.~\cite{mrw1}.

To summarise, the absolute normalisation of the DENZ04 DB$_-$ appears to be in good agreement with our ZUAS12 solution, as is the normalisation of the $\pi^+ p$ backward-angle data sets. Large effects in the normalisation of the $\pi^+ p$ data sets have 
been seen at forward and medium scattering angles~\footnote{It must be added that the absolute normalisation is not the only problem of the DENZ04 measurements. When using the ZUAS12 solution as reference, the shapes of $11$ out of the $29$ data sets 
(after all the outliers are removed) do not pass the test for $\mathrm{p}_{min} \approx 1.24 \cdot 10^{-2}$. The disagreement in shape is very pronounced in the $\pi^+ p$ medium-angle $37.10$ MeV, in the $\pi^- p$ medium/backward-angle $25.80$ MeV, and in 
the $\pi^+ p$ medium-angle $43.30$(rot.) MeV data sets. None of the backward-angle measurements of the DENZ04 DB$_+$ shows any inconsistency in shape when compared with the ZUAS12 solution.}.

The hadronic part of the $\pi N$ interaction is dominant in backward scattering. The importance of the electromagnetic (em) contributions increases with decreasing scattering angle, finally culminating in the Coulomb peak which governs the very-forward 
scattering. The conclusion we drew from the analysis of the DENZ04 DB$_-$ is that the ETH model, along with the optimal values of the model parameters (as obtained in the ZUAS12 solution) and the known em contributions, accounts for the normalisation of the 
data successfully. The conclusion we drew from the analysis of the DENZ04 DB$_+$ is that the hadronic part of the interaction, as deduced on the basis of the ZUAS12 solution, is successful in reproducing the normalisation of the experimental data in the 
backward direction. Due to the fact that the problems lie with the scale factors obtained at forward and medium scattering-angle values, the truncated DENZ04 DB$_+$ seems to indicate modifications in the em part of the $\pi N$ interaction. However, the em 
parts of the two elastic-scattering reactions are intimately connected; one cannot modify one and leave the other intact. (This comment also applies to the hadronic part of the amplitude when involving the ETH model in the fits. The two elastic-scattering 
reactions are linked via the crossing symmetry which the model obeys.) Additionally, the Physics of the Coulomb peak has been established since a very long time.

In view of these results, it is now understood why the fit of the ETH model to the DENZ04 data drifts. As there are no adjustable parameters in the em part of the $\pi N$ interaction, the adjustable hadronic part attempts to compensate for unexpected features 
in the data in a kinematical region in which the sensitivity of the DCS to the hadronic part of the interaction is expected to be low.

Upon inspection of the results of the single-energy phase-shift solution obtained from the DENZ04 data (see Table $6.1$ of Ref.~\cite{denz} and Fig.~4 of the main publication of the CHAOS Collaboration \cite{chaos}), one cannot but feel uneasy about these 
values. Of course, it \emph{is} true that the single-energy phase-shift solutions are not expected to show the smoothness of the results obtained when the energy dependence of the phase shifts is modelled via appropriate functions 
and experimental data, taken at more than one energy, are fitted to, yet the value of the P31 phase shift (i.e., $+0.65^\circ$, no uncertainty has been quoted) in Table $6.1$ of Denz's dissertation, at $19.90$ MeV, is wrong by about $0.9^\circ$; the \emph{largest} 
of the p-wave phase shifts (P33) is itself about $1^\circ$ at that energy! At $20$ MeV, the SAID result \cite{abws} for P31 ($-0.22^\circ$) is almost identical to the value we had obtained in Ref.~\cite{mrw1}. This discrepancy alone provides good reason for 
the thorough re-examination of the results (at least at $19.90$ MeV) of Refs.~\cite{chaos,denz}.

The DENZ04 data cover very low $T$ values, a `corner' of the phase space in which the parameterisations of the K-matrix elements, which we have been using in our analyses for almost two decades, should have worked best. It seems puzzling to be able 
to successfully analyse all the rest of the $\pi N$ data (taken by different groups, with different detectors, at different meson-factory facilities and times) up to $100$ MeV, but be unable to obtain any meaningful results from the DENZ04 measurements, 
which extend only up to $43.30$ MeV. Given the characteristics of the DENZ04 data, there is no room for questioning the theoretical background on which this work relies; we strongly believe that we should have been able to obtain meaningful results from 
these data using both our K-matrix parameterisations, as well as the ETH model.

\section{\label{sec:Discussion}Discussion and Summary}

This paper presents the results of an analysis of the DENZ04 \cite{chaos,denz} low-energy $\pi^\pm p$ differential cross sections. Given the size of the data acquired in this experiment (a total of $546$ data points), the self-consistency of these measurements 
must be addressed prior to their inclusion into our database, which we have carefully established and analysed in our last two phase-shift analyses (PSA) of Refs.~\cite{mworg,mrw1}.

The DENZ04 data were analysed as if they comprised the entire $\pi^\pm p$ elastic-scattering database at low energies, by following the method set forth in Ref.~\cite{mworg}. The analysis of the DENZ04 $\pi^+ p$ measurements on the basis of standard 
low-energy parameterisations of the $s$- and $p$-wave $K$-matrix elements led to the identification of eight outliers in a total of $275$ data points (see Table \ref{tab:DBpi+p}), whereas that of DENZ04 $\pi^- p$ elastic-scattering measurements to the 
removal of three out of a total of $271$ data points (see Table \ref{tab:DBpi-p}).

We subsequently subjected the truncated DENZ04 combined $\pi^\pm p$ elastic-scattering databases into a common optimisation scheme, using the ETH model \cite{glmbg}. The ability of the model to account for the low-energy $\pi N$ interaction has been 
demonstrated during the past two decades of research in this field. To our surprise, the optimisation failed to yield reasonable values for the model parameters. The phase-shift solution, extracted from the fit to the DENZ04 data, is far from the results 
of Refs.~\cite{mworg,mrw1,abws}.

We next tried to trace the origin of these problems by investigating the reproduction of the DENZ04 data on the basis of the results of our recent PSA \cite{mrw1}. We found out that the absolute normalisation of the DENZ04 $\pi^- p$ elastic-scattering 
data is in good agreement with our ZUAS12 solution, as is the normalisation of the DENZ04 $\pi^+ p$ data sets at backward angles. On the other hand, large effects in the normalisation of the DENZ04 $\pi^+ p$ data sets have been seen at forward and medium 
scattering angles, i.e., in the region where the electromagnetic (em) effects are important. Therefore, the DENZ04 data seem to suggest modifications of the em part of the $\pi^+ p$ reaction, whereas the DENZ04 $\pi^- p$ elastic-scattering data are 
compatible with the em part as it currently stands. Given the relation between the em amplitudes for the two reactions, the two aforementioned suggestions are mutually incompatible.

An explanation of the failure of the model to account for the DENZ04 data has been advanced on the basis of the interpretation of Figs.~\ref{fig:sfpip} and \ref{fig:sfpim}. Given that there are no adjustable parameters in the em part of the $\pi N$ interaction, 
the adjustable hadronic part attempts to compensate for unexpected features in the data. In an effort to model large differences in a region where the sensitivity of the DCS to the hadronic part of the interaction is expected to be low, the model parameters drift 
away from their `established' values, to unreasonable ones.

Given all these problems, we are currently unable to include the experimental data of Refs.~\cite{chaos,denz} in our database. Additionally, we would like to remark that the use of these data in low-energy PSAs will surely lead to bias. We hope that the 
findings of the present work will be helpful in the future analyses of the $\pi N$ data.


\begin{ack}
We acknowledge helpful discussions with G.J. Wagner on the acquisition and treatment of the experimental data of the CHAOS Collaboration. We thank G.R. Smith and I.I. Strakovsky for their remarks. Finally, we acknowledge the exchange of interesting ideas with 
W.S. Woolcock(deceased) on a number of issues connected with the present work.

\end{ack}

\newpage
\begin{table}[h!]
{\bf \caption{\label{tab:Pi+p}}}The list of outliers in the DENZ04 $\pi^+ p$ database. The rows represent steps in the outlier-identification/elimination process. The columns indicate: the $\chi^2$ value, the number of degrees of 
freedom NDF in the fit, and the worst data point at that step; the worst data point was then removed and the fit to the remaining data was made. No data can be marked for removal at step $9$. The worst data point is identified 
on the basis of the corresponding pion laboratory kinetic energy $T$ (in MeV) and the centre-of-mass scattering angle $\theta$. The presence of an angular interval at step $6$ indicates that the corresponding data set (i.e., the 
backward-angle data set at $25.80$ MeV) was freely floated in all subsequent fits.
\vspace{0.2cm}
\begin{center}
\begin{tabular}{|c|c|c|l|}
\hline
Step & $\chi^2$ & NDF & Worst data point ($T$, $\theta$) \\
\hline
$1$ & $486.9$ & $268$ & $25.80$, $165.78^\circ$ \\
$2$ & $474.0$ & $267$ & $37.10$, $93.16^\circ$ \\
$3$ & $463.2$ & $266$ & $19.90$, $20.35^\circ$ \\
$4$ & $450.7$ & $265$ & $37.10$, $54.59^\circ$ \\
$5$ & $441.3$ & $264$ & $19.90$, $84.16^\circ$ \\
$6$ & $431.1$ & $263$ & $25.80$, $150.48 - 167.46^\circ$ \\
$7$ & $421.6$ & $262$ & $19.90$, $42.75^\circ$ \\
$8$ & $411.9$ & $261$ & $37.10$, $169.23^\circ$ \\
$9$ & $401.2$ & $260$ & \\
\hline
\end{tabular}
\end{center}
\end{table}

\newpage
\begin{table}
{\bf \caption{\label{tab:DBpi+p}}}The data sets comprising the truncated DENZ04 $\pi^+ p$ database, the pion laboratory kinetic energy $T$ (in MeV), the corresponding angular interval ($\theta$) of the data set (f: forward, m: medium, b: backward), 
the number of degrees of freedom (NDF)$_j$ for each data set, the scale factor $z_j$ which minimises $\chi_j^2$, the values of $(\chi_j^2)_{min}$, and the p-value of the fit for each data set. The numbers of this table correspond to the final fit 
to the data using the $K$-matrix parameterisations (see Section \ref{sec:K-Matrix_pi+p}).
\vspace{0.2cm}
\begin{center}
\begin{tabular}{|c|c|c|c|c|l|}
\hline
$T$, $\theta$ & (NDF)$_j$ & $z_j$ & $(\chi_j^2)_{min}$ & p-value & Comments \\
\hline
$19.90$, f & $5$ & $1.0591$ & $9.3280$ & $0.0967$ & $20.35^\circ$ removed \\
$19.90$, m & $25$ & $0.9378$ & $40.8771$ & $0.0236$ & $42.75$, $84.16^\circ$ removed \\
$25.80$, f & $5$ & $1.0359$ & $12.7392$ & $0.0260$ & \\
$25.80$, m & $27$ & $1.0739$ & $33.0017$ & $0.1970$ & \\
$25.80$, b & $9$ & $0.8150$ & $16.3025$ & $0.0608$ & $165.78^\circ$ removed, freely floated \\
$32.00$, f & $5$ & $1.0299$ & $6.3159$ & $0.2767$ & \\
$32.00$, m & $28$ & $1.0595$ & $44.4274$ & $0.0252$ & \\
$32.00$, b & $13$ & $1.0149$ & $17.9694$ & $0.1587$ & \\
$37.10$, f & $8$ & $0.8957$ & $4.8626$ & $0.7722$ & \\
$37.10$, m & $26$ & $0.8824$ & $41.6963$ & $0.0264$ & $54.59$, $93.16^\circ$ removed \\
$37.10$, b & $12$ & $0.8940$ & $16.2488$ & $0.1801$ & $169.23^\circ$ removed \\
$43.30$, f & $12$ & $0.9905$ & $15.9922$ & $0.1916$ & \\
$43.30$, m & $28$ & $0.9771$ & $37.8442$ & $0.1014$ & \\
$43.30$, b & $13$ & $0.9498$ & $25.6257$ & $0.0191$ & \\
$43.30$(rot.), f & $12$ & $1.0214$ & $23.2806$ & $0.0254$ & \\
$43.30$(rot.), m & $27$ & $0.9934$ & $42.8706$ & $0.0270$ & \\
$43.30$(rot.), b & $12$ & $0.9512$ & $11.8162$ & $0.4606$ & \\
\hline
\end{tabular}
\end{center}
\end{table}

\newpage
\begin{table}[h!]
{\bf \caption{\label{tab:Pi-p}}}The equivalent of Table \ref{tab:Pi+p} for the truncated DENZ04 $\pi^- p$ elastic-scattering database.
\vspace{0.2cm}
\begin{center}
\begin{tabular}{|c|c|c|l|}
\hline
Step & $\chi^2$ & NDF & Worst data point ($T$, $\theta$) \\
\hline
$1$ & $388.2$ & $264$ & $25.80$, $11.08^\circ$ \\
$2$ & $371.5$ & $263$ & $43.30$, $152.55^\circ$ \\
$3$ & $358.9$ & $262$ & $25.80$, $76.00^\circ$ \\
$4$ & $350.2$ & $261$ & \\
\hline
\end{tabular}
\end{center}
\end{table}

\newpage
\begin{table}
{\bf \caption{\label{tab:DBpi-p}}}The equivalent of Table \ref{tab:DBpi+p} for the truncated DENZ04 $\pi^- p$ elastic-scattering database; m/b in the corresponding angular interval ($\theta$) of the data set indicates combined medium and backward angles. 
The numbers of this table correspond to the final fit to the data using the $K$-matrix parameterisations (see Section \ref{sec:K-Matrix_pi-p}).
\vspace{0.2cm}
\begin{center}
\begin{tabular}{|c|c|c|c|c|l|}
\hline
$T$, $\theta$ & (NDF)$_j$ & $z_j$ & $(\chi_j^2)_{min}$ & p-value & Comments \\
\hline
$19.90$, f & $6$ & $1.0075$ & $3.6609$ & $0.7225$ & \\
$19.90$, m & $25$ & $0.9959$ & $18.7750$ & $0.8078$ & \\
$25.80$, f & $6$ & $1.0186$ & $15.1307$ & $0.0193$ & $11.08^\circ$ removed \\
$25.80$, m/b & $37$ & $1.0074$ & $54.0761$ & $0.0346$ & $76.00^\circ$ removed \\
$32.00$, f & $5$ & $0.9525$ & $5.6430$ & $0.3425$ & \\
$32.00$, m/b & $40$ & $0.9918$ & $38.8705$ & $0.5210$ & \\
$37.10$, f & $9$ & $0.9513$ & $8.7386$ & $0.4617$ & \\
$37.10$, m/b & $41$ & $0.9620$ & $59.0529$ & $0.0336$ & \\
$43.30$, f & $12$ & $1.0554$ & $23.3763$ & $0.0247$ & \\
$43.30$, m/b & $38$ & $1.0861$ & $52.6021$ & $0.0579$ & $152.55^\circ$ removed \\
$43.30$(rot.), f & $12$ & $1.1054$ & $21.0420$ & $0.0498$ & \\
$43.30$(rot.), m/b & $37$ & $1.1103$ & $49.2107$ & $0.0864$ & \\
\hline
\end{tabular}
\end{center}
\end{table}

\newpage
\begin{table}
{\bf \caption{\label{tab:Reproduction1}}}The reproduction of the DENZ04 data sets using the ZUAS12 solution as reference (i.e., yielding the `theoretical values' $y_{ij}^{th}$ in Eq.~(1) of Ref.~\cite{mrw1}). The 
columns represent: the pion laboratory kinetic energy $T$ (in MeV), the number of degrees of freedom (NDF)$_j$ for each data set, the reported normalisation uncertainty \cite{chaos}, the scale factor $z_j$ minimising 
$\chi_j^2$ with its uncertainty $\Delta z_j$ (combining in quadrature $\delta z_j$ and the statistical uncertainty derived from Eq.~(2) of Ref.~\cite{mrw1}), the scale factor for free floating $\hat{z}_j$ with 
its uncertainty $\Delta \hat{z}_j$, the minimal $\chi_j^2$ value of the reproduction, the part of $(\chi_j^2)_{min}$ which corresponds to the statistical fluctuation in the data, and the part of $(\chi_j^2)_{min}$ 
which corresponds to the scaling of each data set as a whole. All definitions have been given in Section 2.2 of Ref.~\cite{mrw1}. The table corresponds to the original DENZ04 data sets; all the data, which have been 
obtained at one condition (target configuration, energy), are assumed to comprise one data set in this table. The outliers, detailed in Tables \ref{tab:Pi+p} and \ref{tab:Pi-p}, have \underline{not} been removed.
\vspace{0.2cm}
\begin{center}
\begin{tabular}{|c|c|c|c|c|c|c|c|}
\hline
$T$ & (NDF)$_j$ & $\delta z_j$ & $z_j(\Delta z_j)$ & $\hat{z}_j(\Delta \hat{z}_j)$ & $(\chi_j^2)_{min}$ & $(\chi_j^2)_{st}$ & $(\chi_j^2)_{sc}$ \\
\hline
\multicolumn{8}{|c|}{$\pi^+ p$ scattering} \\
\hline
$19.90$ & $33$ & $0.050$ & $1.077(50)$ & $1.078(50)$ & $118.1$ & $115.7$ & $2.4$ \\
$25.80$ & $43$ & $0.050$ & $1.045(50)$ & $1.046(50)$ & $1208.8$ & $1208.0$ & $0.8$ \\
$32.00$ & $46$ & $0.050$ & $1.129(50)$ & $1.130(50)$ & $340.0$ & $333.3$ & $6.7$ \\
$37.10$ & $49$ & $0.070$ & $0.982(70)$ & $0.982(70)$ & $315.0$ & $314.9$ & $0.1$ \\
$43.30$ & $53$ & $0.070$ & $1.061(70)$ & $1.062(70)$ & $235.6$ & $234.8$ & $0.8$ \\
$43.30$(rot.) & $51$ & $0.070$ & $1.039(70)$ & $1.039(70)$ & $228.5$ & $228.2$ & $0.3$ \\
\hline
\multicolumn{8}{|c|}{$\pi^- p$ elastic scattering} \\
\hline
$19.90$ & $31$ & $0.050$ & $0.963(50)$ & $0.963(50)$ & $54.8$ & $54.3$ & $0.5$ \\
$25.80$ & $45$ & $0.050$ & $1.019(50)$ & $1.019(50)$ & $160.5$ & $160.4$ & $0.1$ \\
$32.00$ & $45$ & $0.050$ & $1.041(50)$ & $1.042(50)$ & $68.6$ & $67.9$ & $0.7$ \\
$37.10$ & $50$ & $0.070$ & $0.999(70)$ & $0.999(70)$ & $84.7$ & $84.7$ & $0.0$ \\
$43.30$ & $51$ & $0.070$ & $1.052(70)$ & $1.052(70)$ & $83.1$ & $82.5$ & $0.6$ \\
$43.30$(rot.) & $49$ & $0.070$ & $1.087(71)$ & $1.088(71)$ & $96.0$ & $94.5$ & $1.6$ \\
\hline
\end{tabular}
\end{center}
\end{table}

\newpage
\begin{table}
{\bf \caption{\label{tab:Reproduction2}}}The equivalent of Table \ref{tab:Reproduction1} in the case that the original $12$ data sets are split in a total of $29$ segments (see Section \ref{sec:General}). The outliers, 
detailed in Tables \ref{tab:Pi+p} and \ref{tab:Pi-p}, have \underline{not} been removed. The corresponding angular interval ($\theta$) of each data set (f: forward, m: medium, b: backward, m/b: combined medium and backward 
angles) is indicated in the first column.
\vspace{0.2cm}
\begin{center}
\begin{tabular}{|c|c|c|c|c|c|c|c|}
\hline
$T$, $\theta$ & (NDF)$_j$ & $\delta z_j$ & $z_j(\Delta z_j)$ & $\hat{z}_j(\Delta \hat{z}_j)$ & $(\chi_j^2)_{min}$ & $(\chi_j^2)_{st}$ & $(\chi_j^2)_{sc}$ \\
\hline
\multicolumn{8}{|c|}{$\pi^+ p$ scattering} \\
\hline
$19.90$, f & $6$ & $0.050$ & $1.084(51)$ & $1.087(51)$ & $24.1$ & $21.1$ & $2.9$ \\
$19.90$, m & $27$ & $0.050$ & $1.070(51)$ & $1.072(51)$ & $95.1$ & $93.1$ & $2.0$ \\
$25.80$, f & $5$ & $0.050$ & $1.063(51)$ & $1.065(51)$ & $21.2$ & $19.6$ & $1.6$ \\
$25.80$, m & $27$ & $0.050$ & $1.208(51)$ & $1.213(51)$ & $62.8$ & $45.0$ & $17.8$ \\
$25.80$, b & $11$ & $0.050$ & $0.824(51)$ & $0.819(51)$ & $42.0$ & $29.2$ & $12.7$ \\
$32.00$, f & $5$ & $0.050$ & $1.157(54)$ & $1.199(56)$ & $19.0$ & $6.5$ & $12.5$ \\
$32.00$, m & $28$ & $0.050$ & $1.193(50)$ & $1.197(50)$ & $73.7$ & $58.5$ & $15.2$ \\
$32.00$, b & $13$ & $0.050$ & $1.023(51)$ & $1.024(51)$ & $19.4$ & $19.1$ & $0.2$ \\
$37.10$, f & $8$ & $0.070$ & $1.076(72)$ & $1.080(72)$ & $9.5$ & $8.2$ & $1.2$ \\
$37.10$, m & $28$ & $0.070$ & $1.008(70)$ & $1.008(70)$ & $110.1$ & $110.1$ & $0.0$ \\
$37.10$, b & $13$ & $0.070$ & $0.912(70)$ & $0.911(70)$ & $27.3$ & $25.7$ & $1.6$ \\
$43.30$, f & $12$ & $0.070$ & $1.225(77)$ & $1.314(83)$ & $28.8$ & $14.4$ & $14.4$ \\
$43.30$, m & $28$ & $0.070$ & $1.130(71)$ & $1.132(71)$ & $45.0$ & $41.5$ & $3.5$ \\
$43.30$, b & $13$ & $0.070$ & $0.983(71)$ & $0.983(71)$ & $23.4$ & $23.3$ & $0.1$ \\
$43.30$(rot.), f & $12$ & $0.070$ & $1.265(77)$ & $1.360(82)$ & $48.7$ & $29.2$ & $19.5$ \\
$43.30$(rot.), m & $27$ & $0.070$ & $1.063(70)$ & $1.064(70)$ & $87.3$ & $86.4$ & $0.8$ \\
$43.30$(rot.), b & $12$ & $0.070$ & $0.984(71)$ & $0.984(71)$ & $10.2$ & $10.2$ & $0.1$ \\
\hline
\multicolumn{8}{|c|}{$\pi^- p$ elastic scattering} \\
\hline
$19.90$, f & $6$ & $0.050$ & $1.004(51)$ & $1.005(51)$ & $3.6$ & $3.6$ & $0.0$ \\
$19.90$, m & $25$ & $0.050$ & $0.940(51)$ & $0.939(51)$ & $31.0$ & $29.5$ & $1.5$ \\
$25.80$, f & $7$ & $0.050$ & $1.011(50)$ & $1.011(50)$ & $38.5$ & $38.5$ & $0.0$ \\
$25.80$, m/b & $38$ & $0.050$ & $1.026(50)$ & $1.027(50)$ & $118.9$ & $118.7$ & $0.3$ \\
$32.00$, f & $5$ & $0.050$ & $0.992(52)$ & $0.991(52)$ & $4.4$ & $4.4$ & $0.0$ \\
\hline
\end{tabular}
\end{center}
\end{table}

\newpage
\begin{table*}
{\bf Table 6 continued}
\vspace{0.2cm}
\begin{center}
\begin{tabular}{|c|c|c|c|c|c|c|c|}
\hline
$T$, $\theta$ & (NDF)$_j$ & $\delta z_j$ & $z_j(\Delta z_j)$ & $\hat{z}_j(\Delta \hat{z}_j)$ & $(\chi_j^2)_{min}$ & $(\chi_j^2)_{st}$ & $(\chi_j^2)_{sc}$ \\
\hline
$32.00$, m/b & $40$ & $0.050$ & $1.050(50)$ & $1.050(50)$ & $49.0$ & $48.0$ & $1.0$ \\
$37.10$, f & $9$ & $0.070$ & $0.986(71)$ & $0.985(71)$ & $7.8$ & $7.8$ & $0.0$ \\
$37.10$, m/b & $41$ & $0.070$ & $1.002(70)$ & $1.002(70)$ & $75.4$ & $75.4$ & $0.0$ \\
$43.30$, f & $12$ & $0.070$ & $1.052(71)$ & $1.054(71)$ & $23.4$ & $22.8$ & $0.6$ \\
$43.30$, m/b & $39$ & $0.070$ & $1.051(71)$ & $1.052(71)$ & $60.2$ & $59.7$ & $0.5$ \\
$43.30$(rot.), f & $12$ & $0.070$ & $1.102(71)$ & $1.106(71)$ & $21.2$ & $19.0$ & $2.2$ \\
$43.30$(rot.), m/b & $37$ & $0.070$ & $1.066(71)$ & $1.069(72)$ & $73.1$ & $72.2$ & $0.9$ \\
\hline
\end{tabular}
\end{center}
\end{table*}

\newpage
\begin{table}
{\bf \caption{\label{tab:Reproduction3}}}The equivalent of Table \ref{tab:Reproduction2} in the case that the outliers, detailed in Tables \ref{tab:Pi+p} and \ref{tab:Pi-p}, have been removed.
\vspace{0.2cm}
\begin{center}
\begin{tabular}{|c|c|c|c|c|c|c|c|}
\hline
$T$, $\theta$ & (NDF)$_j$ & $\delta z_j$ & $z_j(\Delta z_j)$ & $\hat{z}_j(\Delta \hat{z}_j)$ & $(\chi_j^2)_{min}$ & $(\chi_j^2)_{st}$ & $(\chi_j^2)_{sc}$ \\
\hline
\multicolumn{8}{|c|}{$\pi^+ p$ scattering} \\
\hline
$19.90$, f & $5$ & $0.050$ & $1.094(51)$ & $1.098(51)$ & $14.8$ & $11.1$ & $3.7$ \\
$19.90$, m & $25$ & $0.050$ & $1.070(51)$ & $1.072(51)$ & $66.0$ & $64.0$ & $2.0$ \\
$25.80$, f & $5$ & $0.050$ & $1.063(51)$ & $1.065(51)$ & $21.2$ & $19.6$ & $1.6$ \\
$25.80$, m & $27$ & $0.050$ & $1.208(51)$ & $1.213(51)$ & $62.8$ & $45.0$ & $17.8$ \\
$25.80$, b & $9$ & $0.050$ & $0.830(51)$ & $0.830(51)$ & $17.3$ & $17.3$ & $0.0$ \\
$32.00$, f & $5$ & $0.050$ & $1.157(54)$ & $1.199(56)$ & $19.0$ & $6.5$ & $12.5$ \\
$32.00$, m & $28$ & $0.050$ & $1.193(50)$ & $1.197(50)$ & $73.7$ & $58.5$ & $15.2$ \\
$32.00$, b & $13$ & $0.050$ & $1.023(51)$ & $1.024(51)$ & $19.4$ & $19.1$ & $0.2$ \\
$37.10$, f & $8$ & $0.070$ & $1.076(72)$ & $1.080(72)$ & $9.5$ & $8.2$ & $1.2$ \\
$37.10$, m & $26$ & $0.070$ & $1.009(70)$ & $1.009(70)$ & $101.6$ & $101.6$ & $0.0$ \\
$37.10$, b & $12$ & $0.070$ & $0.907(70)$ & $0.906(70)$ & $17.3$ & $15.5$ & $1.8$ \\
$43.30$, f & $12$ & $0.070$ & $1.225(77)$ & $1.314(83)$ & $28.8$ & $14.4$ & $14.4$ \\
$43.30$, m & $28$ & $0.070$ & $1.130(71)$ & $1.132(71)$ & $45.0$ & $41.5$ & $3.5$ \\
$43.30$, b & $13$ & $0.070$ & $0.983(71)$ & $0.983(71)$ & $23.4$ & $23.3$ & $0.1$ \\
$43.30$(rot.), f & $12$ & $0.070$ & $1.265(77)$ & $1.360(82)$ & $48.7$ & $29.2$ & $19.5$ \\
$43.30$(rot.), m & $27$ & $0.070$ & $1.063(70)$ & $1.064(70)$ & $87.3$ & $86.4$ & $0.8$ \\
$43.30$(rot.), b & $12$ & $0.070$ & $0.984(71)$ & $0.984(71)$ & $10.2$ & $10.2$ & $0.1$ \\
\hline
\multicolumn{8}{|c|}{$\pi^- p$ elastic scattering} \\
\hline
$19.90$, f & $6$ & $0.050$ & $1.004(51)$ & $1.005(51)$ & $3.6$ & $3.6$ & $0.0$ \\
$19.90$, m & $25$ & $0.050$ & $0.940(51)$ & $0.939(51)$ & $31.0$ & $29.5$ & $1.5$ \\
$25.80$, f & $6$ & $0.050$ & $1.029(51)$ & $1.030(51)$ & $19.1$ & $18.7$ & $0.3$ \\
$25.80$, m/b & $37$ & $0.050$ & $1.023(50)$ & $1.024(50)$ & $109.9$ & $109.7$ & $0.2$ \\
$32.00$, f & $5$ & $0.050$ & $0.992(52)$ & $0.991(52)$ & $4.4$ & $4.4$ & $0.0$ \\
\hline
\end{tabular}
\end{center}
\end{table}

\newpage
\begin{table*}
{\bf Table 7 continued}
\vspace{0.2cm}
\begin{center}
\begin{tabular}{|c|c|c|c|c|c|c|c|}
\hline
$T$, $\theta$ & (NDF)$_j$ & $\delta z_j$ & $z_j(\Delta z_j)$ & $\hat{z}_j(\Delta \hat{z}_j)$ & $(\chi_j^2)_{min}$ & $(\chi_j^2)_{st}$ & $(\chi_j^2)_{sc}$ \\
\hline
$32.00$, m/b & $40$ & $0.050$ & $1.050(50)$ & $1.050(50)$ & $49.0$ & $48.0$ & $1.0$ \\
$37.10$, f & $9$ & $0.070$ & $0.986(71)$ & $0.985(71)$ & $7.8$ & $7.8$ & $0.0$ \\
$37.10$, m/b & $41$ & $0.070$ & $1.002(70)$ & $1.002(70)$ & $75.4$ & $75.4$ & $0.0$ \\
$43.30$, f & $12$ & $0.070$ & $1.052(71)$ & $1.054(71)$ & $23.4$ & $22.8$ & $0.6$ \\
$43.30$, m/b & $38$ & $0.070$ & $1.050(71)$ & $1.051(71)$ & $51.4$ & $50.8$ & $0.5$ \\
$43.30$(rot.), f & $12$ & $0.070$ & $1.102(71)$ & $1.106(71)$ & $21.2$ & $19.0$ & $2.2$ \\
$43.30$(rot.), m/b & $37$ & $0.070$ & $1.066(71)$ & $1.069(72)$ & $73.1$ & $72.2$ & $0.9$ \\
\hline
\end{tabular}
\end{center}
\end{table*}

\clearpage
\begin{figure}
\begin{center}
\includegraphics [width=15.5cm] {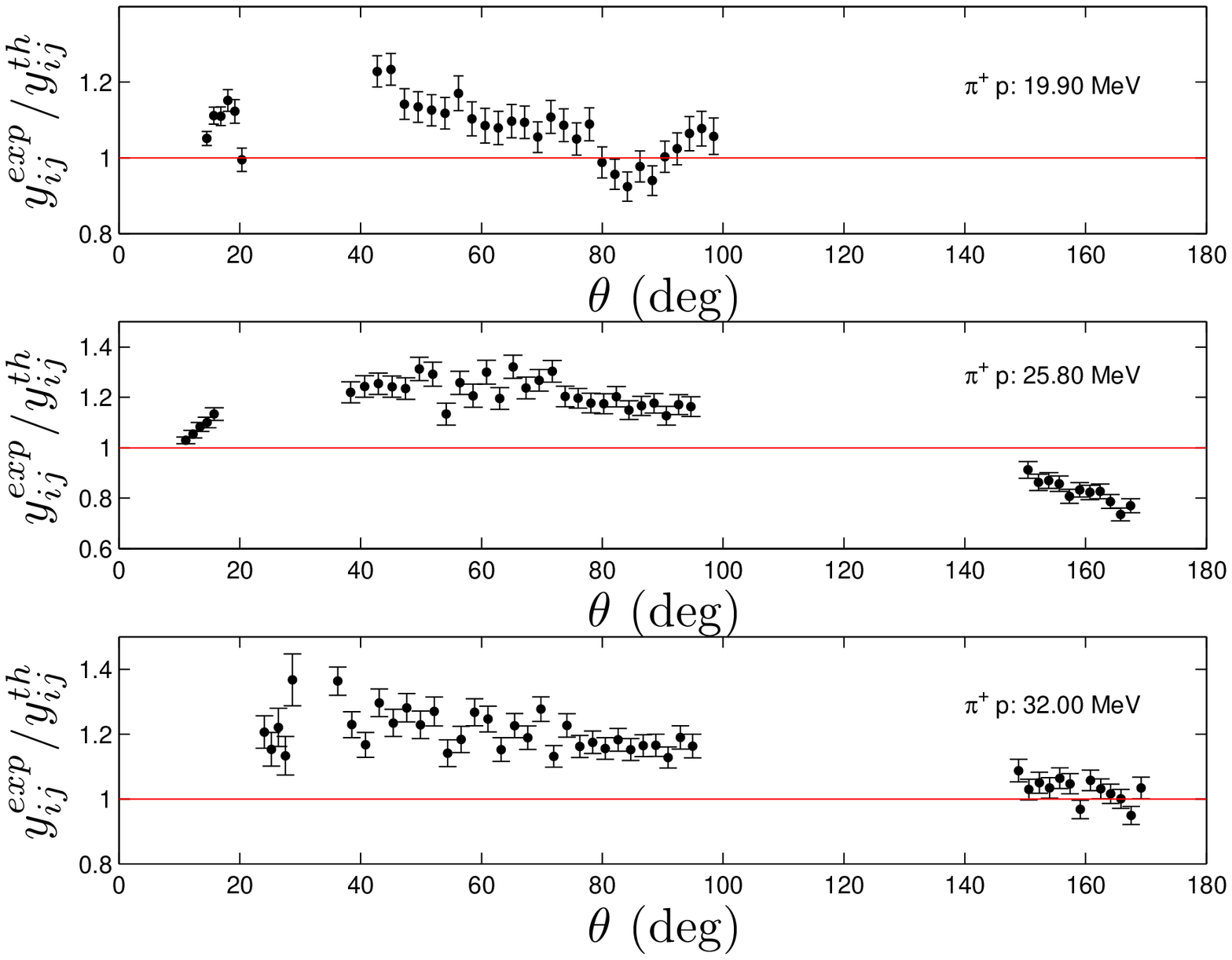}
\end{center}
\end{figure}

\clearpage
\begin{figure}
\begin{center}
\includegraphics [width=15.5cm] {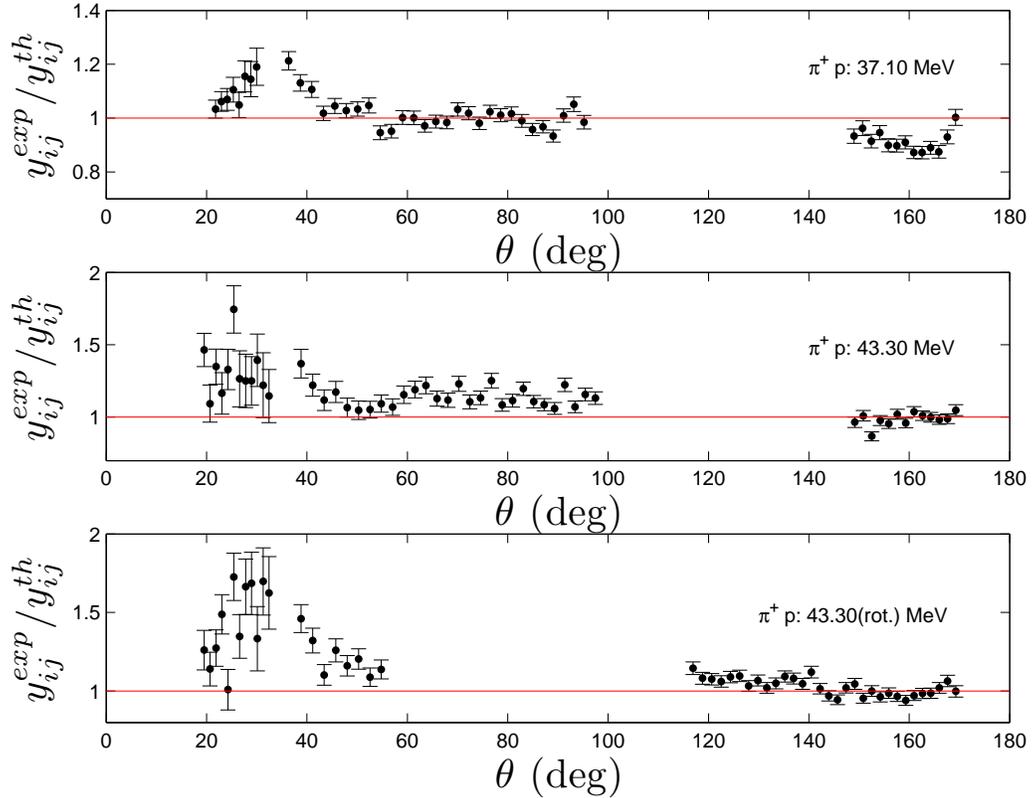}
\caption{\label{fig:PIPPE}The DENZ04 $\pi^+ p$ measurements ($y_{ij}^{exp}$), normalised to the corresponding ZUAS12 predictions ($y_{ij}^{th}$); the outliers detailed in Table \ref{tab:Pi+p} are 
also contained. The normalisation uncertainties of the DENZ04 data sets (see Section \ref{sec:General}) are not shown. The statistical uncertainties of the ZUAS12 predictions (below $5 \%$ in all 
cases, for the energies considered herein) are also not shown.}
\end{center}
\end{figure}

\clearpage
\begin{figure}
\begin{center}
\includegraphics [width=15.5cm] {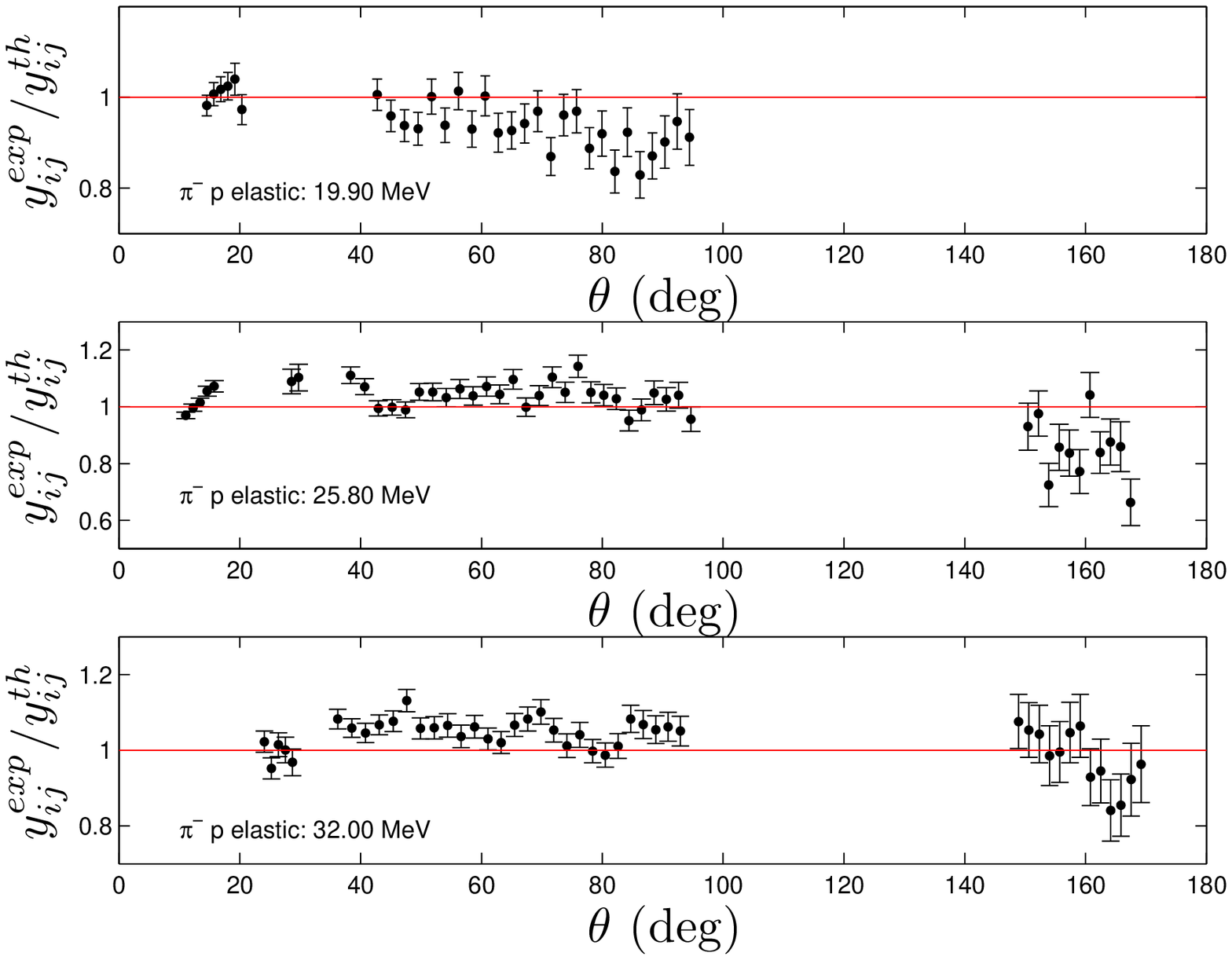}
\end{center}
\end{figure}

\clearpage
\begin{figure}
\begin{center}
\includegraphics [width=15.5cm] {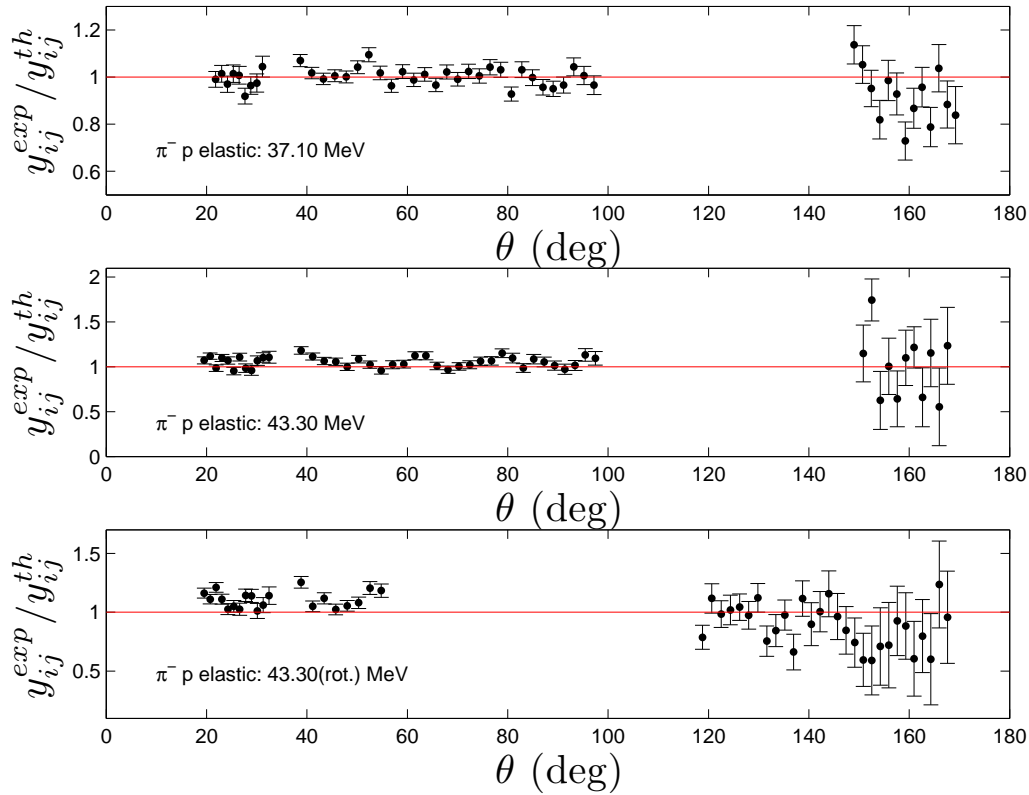}
\caption{\label{fig:PIMPE}The DENZ04 $\pi^- p$ elastic-scattering measurements ($y_{ij}^{exp}$), normalised to the corresponding ZUAS12 predictions ($y_{ij}^{th}$); the outliers detailed in Table 
\ref{tab:Pi-p} are also contained. The normalisation uncertainties of the DENZ04 data sets (see Section \ref{sec:General}) are not shown. The statistical uncertainties of the ZUAS12 predictions 
(typically at the few-percent level; between $6$ and about $23 \%$ in the very backward direction, for the energies considered herein) are also not shown.}
\end{center}
\end{figure}

\clearpage
\begin{figure}
\begin{center}
\includegraphics [width=15.5cm] {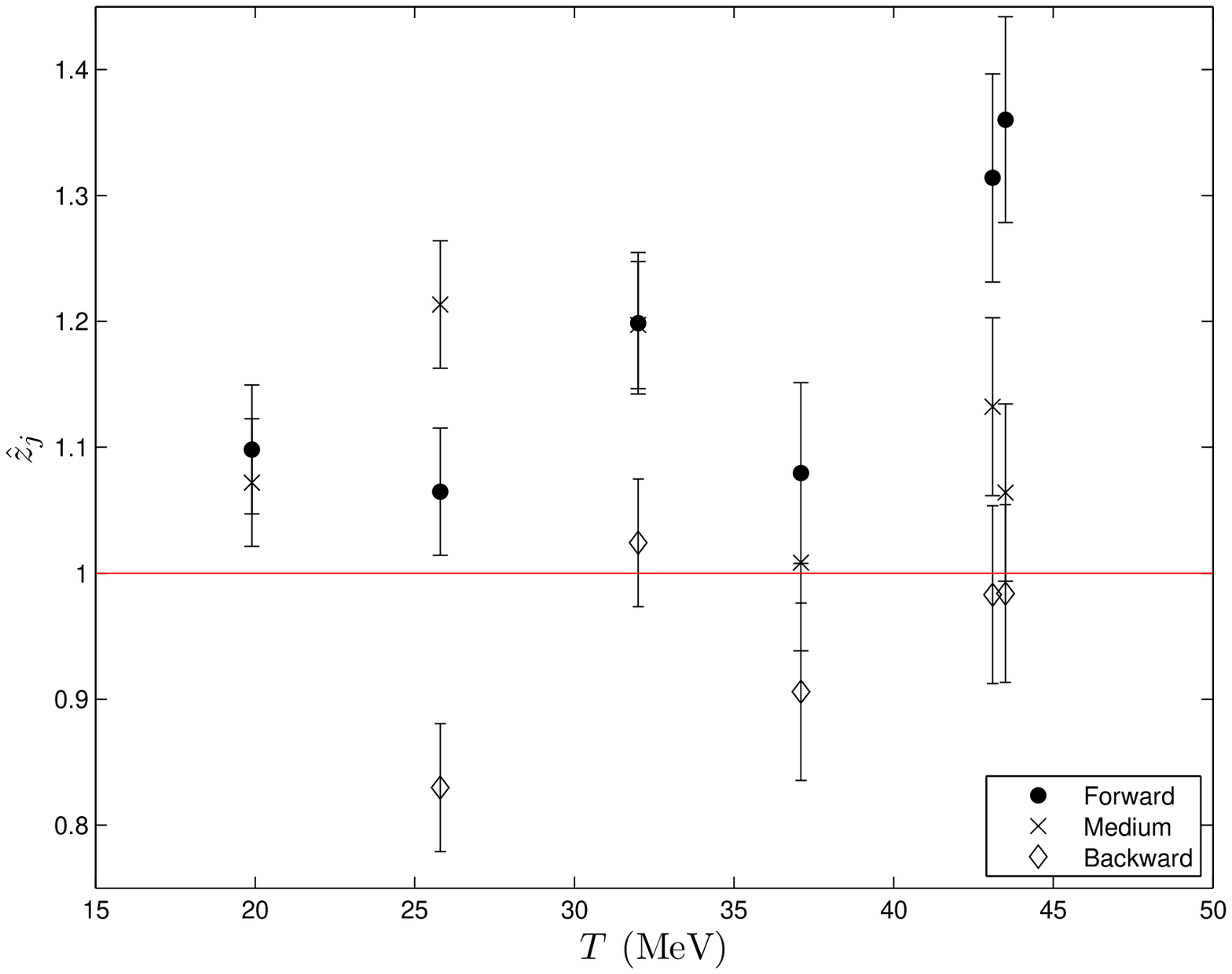}
\caption{\label{fig:sfpip}The scale factors for free floating $\hat{z}_j$ for the DENZ04 $\pi^+ p$ data, obtained on the basis of the ZUAS12 solution. To improve the display, the $\hat{z}_j$ values 
at $43.30$ MeV are shown slightly shifted (horizontally). The labels `forward', `medium', and `backward' indicate the corresponding angular interval of the measurements.}
\end{center}
\end{figure}

\clearpage
\begin{figure}
\begin{center}
\includegraphics [width=15.5cm] {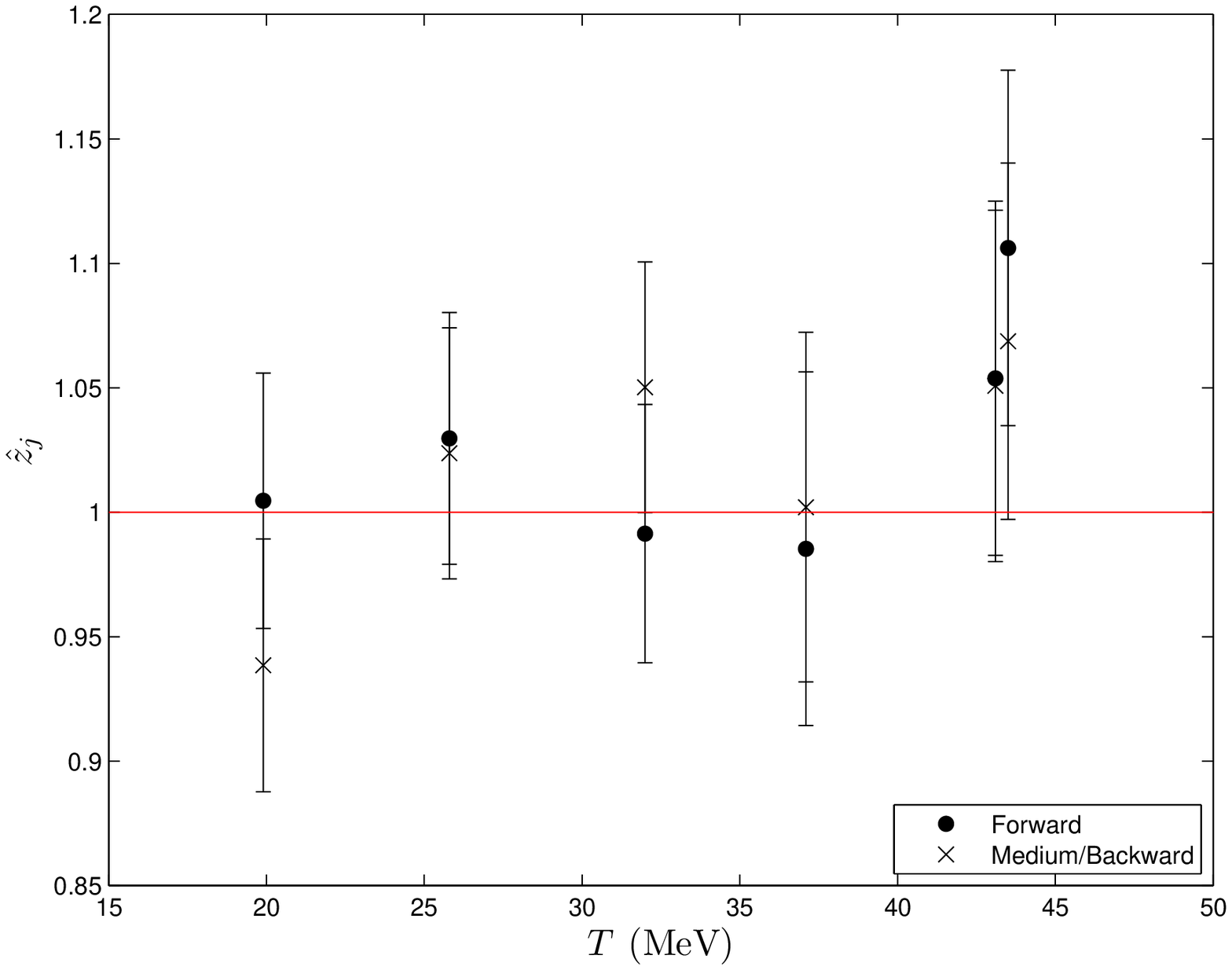}
\caption{\label{fig:sfpim}The scale factors for free floating $\hat{z}_j$ for the DENZ04 $\pi^- p$ elastic-scattering data, obtained on the basis of the ZUAS12 solution. To improve the display, the 
$\hat{z}_j$ values at $43.30$ MeV are shown slightly shifted (horizontally). The labels `forward' and `medium/backward' indicate the corresponding angular interval of the measurements.}
\end{center}
\end{figure}

\end{document}